\documentclass[conference]{IEEEtran}
\IEEEoverridecommandlockouts
\usepackage{cite}
\usepackage{amsmath,amssymb,amsfonts}
\usepackage{algorithmic}
\usepackage{algorithm}
\usepackage{graphicx}
\usepackage{textcomp}
\usepackage{xcolor}
\usepackage{stfloats}

\usepackage{lipsum}
\usepackage{mathtools}
\usepackage{cuted}

\usepackage[nodisplayskipstretch]{setspace}
\usepackage{bm}
\usepackage{subfigure}
\usepackage{amsthm} 

\def\BibTeX{{\rm B\kern-.05em{\sc i\kern-.025em b}\kern-.08em
		T\kern-.1667em\lower.7ex\hbox{E}\kern-.125emX}}
\newcommand{\mv}[1]{\mbox{\boldmath{$ #1 $}}}

\newtheorem{proposition}{Proposition}
\newtheorem{remark}{Remark}

\newtheorem{lemma}{Lemma}

\makeatletter
\def\endthebibliography{%
	\def\@noitemerr{\@latex@warning{Empty `thebibliography' environment}}%
	\endlist
}
\makeatother

\begin{document}
	\title{Near-Field 3D Localization via MIMO Radar: Cram\'er-Rao Bound and Estimator Design \\
	}
	\author{\IEEEauthorblockN{Haocheng~Hua
			and
			Jie~Xu
		}
		\IEEEauthorblockA{School of Science and Engineering and Future Network of Intelligence Institute,\\ The Chinese University of Hong Kong (Shenzhen), Shenzhen, China\\
				Email: haochenghua@link.cuhk.edu.cn,
				~xujie@cuhk.edu.cn}
			\thanks{Jie Xu is the corresponding author.}
		}
		
		\maketitle
		
		\begin{abstract}
			Future sixth-generation (6G) networks are envisioned to provide both sensing and communications functionalities by using densely deployed base stations (BSs) with massive antennas operating in millimeter wave (mmWave) and terahertz (THz). Due to the large number of antennas and the high frequency band, the sensing and communications are expected to be implemented within the near-field region, thus making the conventional designs based on the far-field channel models inapplicable. This paper studies a near-field multiple-input-multiple-output (MIMO) radar sensing system, in which the transceivers with massive antennas aim to localize multiple near-field targets in the three-dimensional (3D) space. In particular, we adopt a general wavefront propagation model by considering the exact spherical wavefront with both channel phase and amplitude variations over different antennas. Besides, we consider the general transmit signal waveforms and also consider the unknown cluttered environments. Under this setup, the unknown parameters to estimate include the 3D coordinates and the complex reflection coefficients of the multiple targets, as well as the noise and interference covariance matrix. Accordingly, we derive the Fisher information matrix (FIM) corresponding to the 3D coordinates and the complex reflection
			coefficients of the targets and accordingly obtain the CRB for estimating the 3D coordinates. This provides a performance bound for 3D near-field target localization. Next, to facilitate practical localization, we propose an efficient estimation algorithm based on the 3D approximate cyclic optimization (3D-ACO), which is obtained following the maximum likelihood (ML) criterion. Finally, numerical results show that considering the exact antenna-varying channel amplitudes achieves more accurate CRB as compared to prior works based on constant channel amplitudes across antennas, especially when the targets are close to the transceivers. It is also shown that the proposed estimator achieves localization performance close to the derived CRB, thus validating its effectiveness in practical implementation. 
		\end{abstract}
		
		
		\section{Introduction} 
		\label{sec:intro}
		Integrated sensing and communication (ISAC) has been recognized as one of the key candidate technologies towards sixth-generation (6G) wireless networks in both academia and industry, in which the radar sensing functionality is integrated into wireless networks for providing new services and enhancing the communication performance \cite{liu2022integrated,hua2023optimal,liu2021cramer}. On the other hand, ultra-dense base station (BS) deployment, millimeter wave (mmWave) and terahertz (THz),	and massive multiple-input multiple-output (MIMO) techniques \cite{heath2016overview} have been widely adopted in wireless networks. As such, the BSs may need to perform the sensing and communications in the near-field region (or Fresnel region \cite{khamidullina2021conditional}), for which the conventional far-field channel model considering the planar wavefront assumption of the electromagnetic (EM) waves is no longer valid or accurate. This thus introduces a paradigm shift from the conventional far-field sensing and communications design to the new near-field design, for which the spherical wavefront should be considered \cite{lu2021does,lu2023near,khamidullina2021conditional}.
		
		This paper focuses on the near-field radar sensing under the spherical wavefront model. In this case, we can use one antenna array to resolve both the angles of arrival (AoAs) and the ranges of the sensing targets \cite{khamidullina2021conditional,huang1991near}. First, by considering uniform linear array (ULA), various prior works considered the near-field target localization in the two-dimensional (2D) plane (see, e.g., \cite{huang1991near}), in which near-field estimators were designed based on one-dimensional (1D) array processing. Furthermore, by adopting uniform planar array (UPA) or other 2D antenna array configurations, recent work \cite{khamidullina2021conditional} considered
		the near-field target localization in the three-dimensional (3D) space, in which 2D array processing is necessary for resolving the exact 3D target coordinates.
		
		Same as \cite{khamidullina2021conditional}, we study the near-field 3D target localization. Under this setup, prior work \cite{khamidullina2021conditional} aimed to localize multiple targets by considering spherical wavefront models with varying phases but constant amplitudes across different antenna elements. Based on this model, the authors in \cite{khamidullina2021conditional} derived the Cram\'er-Rao Bound (CRB) for estimating the target 3D coordinates, which characterizes the variance lower bound by any unbiased estimators \cite{levy2008principles}. However, the analysis in \cite{khamidullina2021conditional} has the following limitations. First, the antenna-independent constant amplitudes were considered in \cite{khamidullina2021conditional}, which, however, cannot capture the amplitude variation over different antennas. This assumption may result in inaccurate CRB analysis, especially when the number of antennas become significantly large and/or the distances between the transceiver and the targets become sufficiently short. Next, the orthogonal waveform transmission and matched filtering reception were considered in \cite{khamidullina2021conditional}. This, however, cannot work well for sensing in ISAC networks, when the BSs need to employ other waveforms to simultaneously performing other tasks such as communications or when the BSs can optimize the transmit waveform based on prior target information in, e.g., target tracking phases \cite{liu2021cramer,hua2022mimojournal}. Furthermore, only white Gaussian noise (WGN) was considered in \cite{khamidullina2021conditional}, which may lead to highly suboptimal performance in unknown cluttered environments. Therefore, to overcome the above limitations, we study the 3D near-field targets localization in this work, by considering more generic transmit waveform, channel propagation, and clutter models. 
		
		In particular, this paper studies a near-field MIMO radar sensing system, in which the transceivers with massive antennas aim to localize multiple near-field targets in the 3D space. We adopt an exact spherical wavefront model by considering the distance-dependent variations in both channel phases and amplitudes over different antennas. We also consider the general transmit signal waveforms and the unknown cluttered environments. Under this setup, the unknown parameters to estimate at the transceivers include the 3D coordinates and the complex reflection coefficients of the multiple targets, as well as the noise and interference covariance matrix. First, we derive the Fisher information matrix (FIM) corresponding to the 3D coordinates and the complex reflection coefficients of the targets and accordingly obtain the CRB, which provides a performance bound for
		3D near-field target localization. Next, to facilitate practical localization, we propose an efficient estimation algorithm based on the 3D approximate cyclic optimization (3D-ACO), which is obtained following the maximum likelihood (ML) criterion.  Finally, we conduct numerical results to validate the necessity of considering the channel amplitude variation and evaluate the performance of the proposed estimator. It is shown that when the targets are close to the transceiver, our considered model with varying channel amplitudes achieves more accurate CRB than that without such consideration. It is also shown that the proposed 3D-ACO estimation algorithm achieves the mean square error (MSE) performance close to the derived CRB, especially in the high signal-to-noise ratio (SNR) regime.

		{\it Notations:} Boldface letters refer to vectors (lower case) or matrices (upper case). For a square matrix $\mv{M}$, ${\operatorname{tr}}(\mv{M})$, $\bm{M}^{-1}$, and $|\bm{M}|$ denote its trace, inverse, and determinant, respectively. For an arbitrary-size matrix $\mv{M}$, $\mathfrak{R}(\bm{M})$, $\mathfrak{I}(\bm{M})$, $\bm{M}^H$, $\bm{M}^*$, $\bm{M}^T$, $\bm{M}[m:n,p:q]$, and $\operatorname{vec}(\bm{M})$ denote its real part, imaginary part, conjugate transpose, conjugate, transpose, the corresponding sub-block matrix with dimension $(n-m+1) \times (q-p+1)$, and its vectorization, respectively. $\odot$ denotes the Hadamard product.
		The distribution of a circularly symmetric complex Gaussian (CSCG) random vector with mean vector $\mv{x}$ and covariance matrix $\mv{\Sigma}$ is denoted by $\mathcal{CN}(\mv{x,\Sigma})$; and $\sim$ stands for ``distributed as''.
		$\mathbb{R}^{x\times y}$ and $\mathbb{C}^{x\times y}$ denote the spaces of real and complex matrices with dimension $x \times y$, respectively. {${\mathbb{E}}\{\cdot\}$} denotes the statistical expectation. $\|\mv{x}\|$ denotes the Euclidean norm of a complex vector $\mv{x}$ and $\operatorname{diag}(\bm{x})$ denotes a diagonal matrix with diagonal elements $\bm{x}$. $\mathrm{j} = \sqrt{-1}$ denotes the imaginary unit.

		\section{System Model}\label{Section_system_model}
		
\begin{figure}[t]
			\centering
			\includegraphics[width=3.1in]{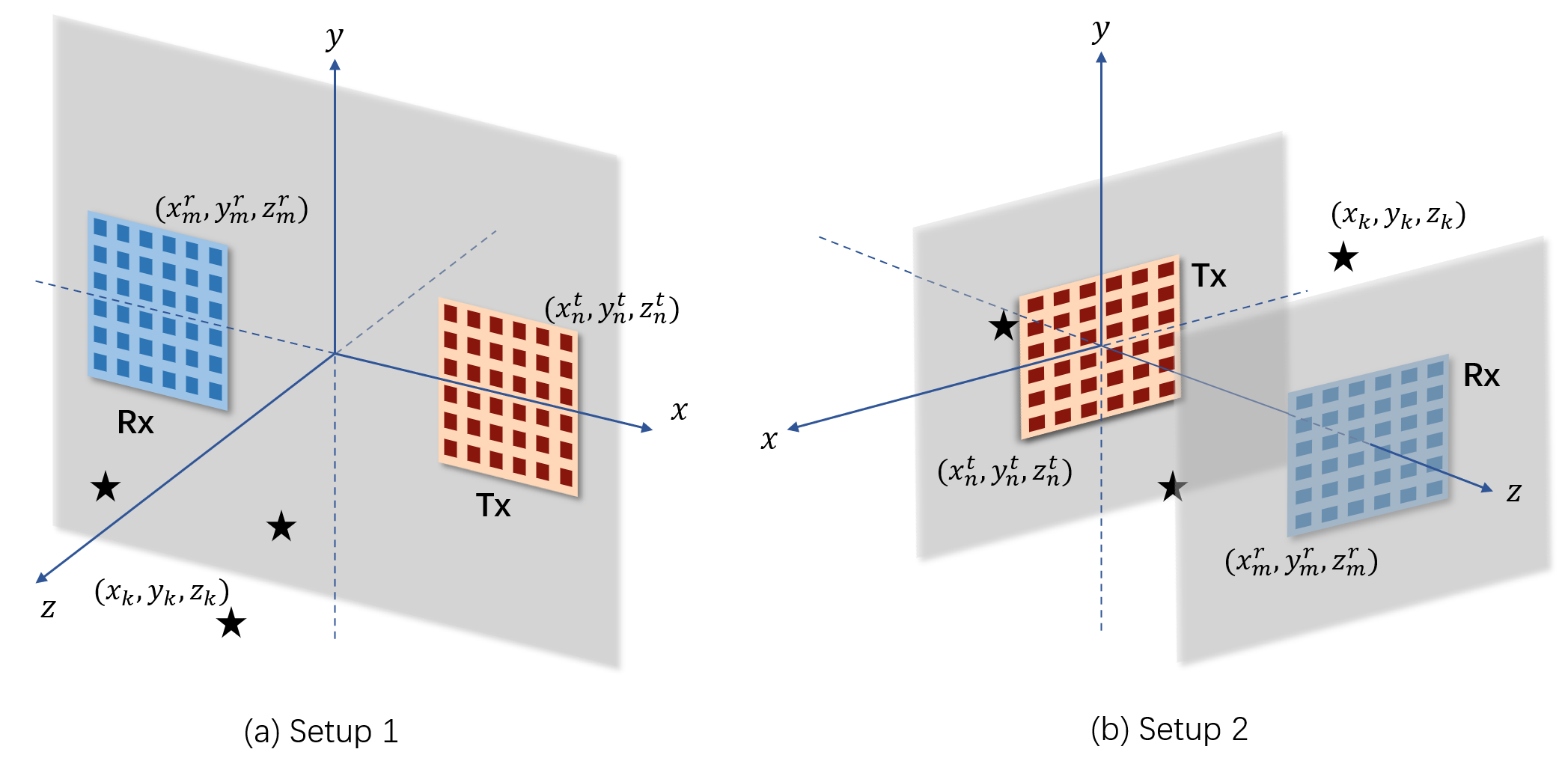}
			\centering
			\caption{MIMO radar operating in near-field region, where black stars denote the targets to be localized: (a) Antenna arrays at Rx and Tx are located in the same plane; (b) Antenna arrays at Rx and Tx face towards each other.}
			\label{fig:sm}
		\end{figure}

		Consider a MIMO radar system equipped with $ N $ antennas at the transmitter (Tx) and $M$ antennas at the receiver (Rx).
		Figs. \ref{fig:sm}(a) and \ref{fig:sm}(b) show two geometry configurations when the Tx and Rx antenna arrays are deployed in the same and different planes, respectively. Notice that  Fig. \ref{fig:sm}(a) may contain the mono-static MIMO radar as a special case when  the Tx and Rx are overlapped with each other.
		Let $\bm{l}^t_{n} = (x^t_n,y^t_n,z^t_n), n \in \{1,2,...,N\}$, and $\bm{l}^r_{m} = (x^r_m,y^r_m,z^r_m), m \in \{1,2,...,M\}$, denote the position of the $n$-th antenna at Tx and that of the $m$-th antenna at Rx, respectively. There exists $K$ sensing targets located in the near-field region of the MIMO radar and the position of the $k$-th target is denoted by $\bm{l}_k = (x_k,y_k,z_k)$. Let $\mathcal{K} \triangleq \{1,...,K\}$ denote the set of targets. 
		Let $\bm{x}_l$ denote the transmit signal at the Tx corresponding to symbol $l \in \{1,...,L\}$ for multi-target localization, and $\bm{X} = \left[\bm{x}_1,...,\bm{x}_L\right] \in \mathbb{C}^{N \times L}$ denote the transmit signals over the $L$ symbols. The sample covariance matrix of the transmitted signals is then given by
		\begin{align}\label{eq:remain_def_Fisher_Rx}
			\bm{R}_X = \frac{1}{L} \bm{X} \bm{X}^H.
		\end{align}
		Under the narrowband assumption, the received signal over the $L$ symbols is described by matrix $\bm{Y} \in \mathbb{C}^{M \times L}$, given by
		\begin{align}\label{equ:Rx_data_matrix}
			\bm{Y} = \sum_{k = 1}^{K} \bm{a}(\bm{l}_k) b_k \bm{v}^T(\bm{l}_k) \bm{X} + \bm{Z},
		\end{align}
		where $\bm{Z} = \left[\bm{z}_1,...,\bm{z}_L\right] \in \mathbb{C}^{M \times L}$ denotes the noise and clutter with each of its column being independent and identically distributed (i.i.d.) CSCG random vectors with zero mean and covariance $\bm{Q} \in \mathbb{C}^{M \times M}$, and $b_k$ denotes the target complex reflection coefficients proportional to the radar-cross-sections (RCS) of the $k$-th target. 
		Here, we consider the unknown cluttered environments, such that $\bm{Q}$ is assumed to be an unknown to be estimated. Furthermore, in (\ref{equ:Rx_data_matrix}),
		$\bm{a}(\bm{l}_k)$ and $\bm{v}(\bm{l}_k)$ denote the steering vectors at Rx and Tx for the $k$-th target, respectively, given by 
		\begin{align}\label{equ:steer_Rx}
			\bm{a}(\bm{l}_k) = [A_1^r(\bm{l}_k)  e^{-\mathrm{j} \nu \|\bm{l}^r_1 - \bm{l}_{k}\|},..., A_{M}^r(\bm{l}_k) e^{-\mathrm{j} \nu \|\bm{l}^r_M - \bm{l}_{k}\|} ]^T,
		\end{align}
		\begin{align}\label{equ:steer_Tx}
			\bm{v}(\bm{l}_k) = [A_1^t(\bm{l}_k)e^{-\mathrm{j} \nu \|\bm{l}^t_1 - \bm{l}_{k}\|},..., A_N^t(\bm{l}_k)e^{-\mathrm{j} \nu \|\bm{l}^t_N - \bm{l}_{k}\|} ]^T,
		\end{align}
		where $\nu = \frac{2 \pi}{\lambda}$ is the wave number of the carrier, $\lambda$ is the wavelength, and $A_m^r(\bm{l}_k) = \frac{\lambda \sqrt{G_r}}{4 \pi \|\bm{l}_k - \bm{l}^r_m\|}$ and $A_n^t(\bm{l}_k) = \frac{\lambda \sqrt{G_t}}{4 \pi \|\bm{l}_k - \bm{l}^t_{n}\|}$ represent the distance-dependent channel amplitude from the $k$-th target to the $m$-th receive antenna and that from the $n$-th transmit antenna to the $k$-th target based on the free space path-loss model. We further assume omnidirectional antennas with unit antenna gains, i.e., $G_r = G_t = 1$. 
		
		For ease of exposition, the received signal $\bm{Y}$ in (\ref{equ:Rx_data_matrix}) can be expressed in a more compact form as
		\begin{align}\label{equ:data_compact_rx}
			\bm{Y} = \bm{A}(\{\bm{l}_k\}) \bm{B} \bm{V}^T(\{\bm{l}_k\}) \bm{X} + \bm{Z},
		\end{align}
		with
		\begin{align}
			\label{equ:A_compact}
			& \bm{A}(\{\bm{l}_k\}) = \left[\bm{a}(\bm{l}_1),\bm{a}(\bm{l}_2),...,\bm{a}(\bm{l}_K)\right] \in \mathbb{C}^{M \times K}, \\
			\label{equ:V_compact}
			& \bm{V}(\{\bm{l}_k\})  = \left[\bm{v}(\bm{l}_1),\bm{v}(\bm{l}_2),...,\bm{v}(\bm{l}_K)\right] \in \mathbb{C}^{N \times K}, \\
			\label{equ:b_complete_RI}
			& \bm{b}  = \left[b_1,...,b_K\right]^T =  \left[b_{\text{R}_1}+ \mathrm{j}b_{\text{I}_1},...,b_{\text{R}_K}+\mathrm{j}b_{\text{I}_K}\right]^T, \\
			\label{equ:b_compact}
			& \qquad \qquad \qquad \bm{B}  = \operatorname{diag}(\bm{b}),
		\end{align}
		where $b_{\text{R}_k}$ and $b_{\text{I}_k}$ denote the real and imaginary parts of $b_k$, respectively. 
		In (\ref{equ:data_compact_rx}), the unknown parameters include the 3D target positions $\{\bm{l}_k\}$, the complex reflection coefficients $\bm{b}$, and the noise and interference covariance matrix $\bm{Q}$. 
		
		\section{Near-Field CRB}\label{sec:nf_CRB}
		
		In this section, we derive the CRB for estimating the 3D coordinates of the multiple near-field targets based on (\ref{equ:data_compact_rx}). 
		Let $\tilde{\bm{\theta}} \in \mathbb{R}^{5K+M^2}$ denote a vector containing all the real unknowns in the target-related parameter vector $\bm{\theta}$, defined as
		\begin{align}\label{eq:theta_interested}
			\nonumber
			\bm{\theta} & = \left[\mathrm{x}_1,...,\mathrm{x}_K,\mathrm{y}_1,...,\mathrm{y}_K,\mathrm{z}_1,...,\mathrm{z}_K,b_{\text{R}_1},..,b_{\text{R}_K},b_{\text{I}_1},...,b_{\text{I}_K}\right]^T \\
			& = \left[\bm{\mathrm{x}},\bm{\mathrm{y}},\bm{\mathrm{z}},\bm{b}_\text{R},\bm{b}_\text{I}\right]^T \in \mathbb{R}^{5K},
		\end{align}
		and the $M^2$ unknown nuisance real parameters in $\bm{Q}$. Let $\bm{y} \triangleq \operatorname{vec}(\bm{Y})$. It follows from (\ref{equ:data_compact_rx}) that
		\begin{align}\label{eq:vec_Y}
			\bm{y} = \left[(\bm{A} \bm{B} \bm{V}^T \bm{x}_1)^T+\bm{z}_1^T,...,(\bm{A} \bm{B} \bm{V}^T \bm{x}_L)^T+\bm{z}_L^T\right]^T,
		\end{align}
		which is a complex Gaussian random vector with mean vector
		\begin{align}\label{equ:Rx_data_mat_DGC}
			\bm{\mathrm{\mu}}(\tilde{\bm{\theta}}) = \left[(\bm{A} \bm{B} \bm{V}^T \bm{x}_1)^T,...,(\bm{A} \bm{B} \bm{V}^T \bm{x}_L)^T\right]^T
		\end{align}
		and covariance matrix
		\begin{align}\label{equ:cov_C}
			\bm{\bm{C}}(\tilde{\bm{\theta}}) = \left[\begin{array}{ccc}
				\bm{Q} & \bm{0} & \bm{0} \\
				\bm{0}&  \ddots & \bm{0}\\
				\bm{0}  &\bm{0} & \bm{Q}  
			\end{array}\right] \in \mathbb{C}^{ML \times ML},
		\end{align}
		i.e., $\bm{y} \sim \mathcal{CN}(\bm{\mu}(\tilde{\bm{\theta}}),\bm{C}(\tilde{\bm{\theta}}))$. Here, (\ref{equ:cov_C}) holds since the observations at different snapshots are uncorrelated with each other. 
		
		First, we obtain the FIM for parameters estimation. Towards this end, we first present the following lemma\cite[B.3.25]{stoica2005spectral}.
		\begin{lemma}\label{lemma:Fisher_deri}
			\emph{Assume that the observations are an $N_{\zeta}$-vector $\bm{y}_{\zeta} \sim \mathcal{CN}(\hat{\bm{\mu}}(\bm{\zeta}), \hat{\bm{C}}(\bm{\zeta}))$, the $(i,j)$-th element of the FIM $\bm{\mathcal{F}}(\bm{\zeta}_i,\bm{\zeta}_j)$, is given as
				\begin{align}\label{equ:fund_Gauss_CRB}
					\nonumber
					\bm{\mathcal{F}}(\bm{\zeta}_i,\bm{\zeta}_j)  = & \operatorname{tr}\left[\hat{\bm{C}}^{-1}(\bm{\zeta}) \frac{\partial \hat{\bm{C}}(\bm{\zeta})}{\partial \bm{\zeta}_i} \hat{\bm{C}}^{-1}(\bm{\zeta}) \frac{\partial \hat{\bm{C}}(\bm{\zeta})}{\partial \bm{\zeta}_j}\right] \\
					& + 2 \mathfrak{R} \left[  (\frac{\partial \hat{\bm{\mu}}(\bm{\zeta})}{\partial \bm{\zeta}_i})^H \hat{\bm{C}}^{-1}(\bm{\zeta}) \frac{\partial \hat{\bm{\mu}}(\bm{\zeta})}{\partial \bm{\zeta}_j}\right].
				\end{align}
			}
		\end{lemma}
		According to Lemma \ref{lemma:Fisher_deri}, we have the following Proposition.
		\begin{proposition}\label{Pro:overall_fisher}
			\emph{  The overall FIM of $\tilde{\bm{\theta}}$, denoted as $\tilde{\bm{\mathrm{F}}} \in \mathbb{R}^{(5K+M^2) \times (5K+M^2)}$, is given as
				\begin{align}\label{eq:overall_fisher}
					\tilde{\bm{\mathrm{F}}} & (\tilde{\bm{\theta}}_i,\tilde{\bm{\theta}}_j)   = L \operatorname{tr}\left[\bm{Q}^{-1} \frac{\partial \bm{Q}}{\partial \tilde{\bm{\theta}}_i} \bm{Q}^{-1} \frac{\partial \bm{Q}}{\partial \tilde{\bm{\theta}}_j}\right] + \\
					\nonumber
					& 2 \mathfrak{R} \operatorname{tr} \left[ (\frac{\partial (\bm{A} \bm{B} \bm{V}^T \bm{X})}{\partial \tilde{\bm{\theta}}_i})^H \bm{Q}^{-1} (\frac{\partial (\bm{A} \bm{B} \bm{V}^T \bm{X})}{\partial \tilde{\bm{\theta}}_j}) \right],
				\end{align}
				where $\tilde{\bm{\mathrm{F}}}(\tilde{\bm{\theta}}_i,\tilde{\bm{\theta}}_j)$ denotes the $(i,j)$-th element of the FIM $\tilde{\bm{\mathrm{F}}}$.
				\begin{proof}
					Please refer to Appendix \ref{Appendix:Proof_Prop_1}.
			\end{proof}}
		\end{proposition}

		Based on Proposition \ref{Pro:overall_fisher}, one can find that
		the overall FIM $\tilde{\bm{\mathrm{F}}}$ is a block diagonal matrix with respect to the target-related unknowns in $\bm{\theta}$ and the nuisance parameters in $\bm{Q}$. Therefore, we can calculate the CRBs of $\bm{\theta}$ and $\bm{Q}$ separately. As we are mainly interested in the CRBs of $\{\bm{\mathrm{l}}_k\}$, we just need to find the FIM $\bm{\mathrm{F}}$ corresponding to $\{\{\bm{\mathrm{l}}_k\}, \bm{b}\}$, i.e., $\bm{\mathrm{F}} = \tilde{\bm{\mathrm{F}}}[1:5K,1:5K]$, which is shown in the following proposition.

		\begin{proposition}\label{Pro:F_deri}
			\emph{The FIM $\bm{\mathrm{F}} \in \mathbb{R}^{5K \times 5K}$ with respect to $\bm{\theta}$ is given as
				\begin{align}\label{eq:fisher_mat_sim}
								2 \left[
					\arraycolsep=0.75pt\def\arraystretch{1.2}
					\begin{array}{ccccc}
						\mathfrak{R}(\bm{\mathrm{F}}_{\bm{xx}}) & \mathfrak{R}(\bm{\mathrm{F}}_{\bm{xy}}) & \mathfrak{R}(\bm{\mathrm{F}}_{\bm{xz}}) & \mathfrak{R}(\bm{\mathrm{F}}_{\bm{xb}}) & -\mathfrak{I}(\bm{\mathrm{F}}_{\bm{xb}}) \\[1pt]
						\mathfrak{R}(\bm{\mathrm{F}}^T_{\bm{xy}}) & \mathfrak{R}(\bm{\mathrm{F}}_{\bm{yy}}) & \mathfrak{R}(\bm{\mathrm{F}}_{\bm{yz}}) & \mathfrak{R}(\bm{\mathrm{F}}_{\bm{yb}}) & -\mathfrak{I}(\bm{\mathrm{F}}_{\bm{yb}}) \\[1pt]
						\mathfrak{R}(\bm{\mathrm{F}}^T_{\bm{xz}}) & \mathfrak{R}(\bm{\mathrm{F}}^T_{\bm{yz}}) & \mathfrak{R}(\bm{\mathrm{F}}_{\bm{zz}}) & \mathfrak{R}(\bm{\mathrm{F}}_{\bm{zb}}) & -\mathfrak{I}(\bm{\mathrm{F}}_{\bm{zb}}) \\[1pt]
						\mathfrak{R}(\bm{\mathrm{F}}^T_{\bm{xb}}) & \mathfrak{R}(\bm{\mathrm{F}}^T_{\bm{yb}}) & \mathfrak{R}(\bm{\mathrm{F}}^T_{\bm{zb}}) & \mathfrak{R}(\bm{\mathrm{F}}_{\bm{bb}}) & -\mathfrak{I}(\bm{\mathrm{F}}_{\bm{bb}}) \\[1pt]
						-\mathfrak{I}(\bm{\mathrm{F}}^T_{\bm{xb}}) & -\mathfrak{I}(\bm{\mathrm{F}}^T_{\bm{yb}}) & -\mathfrak{I}(\bm{\mathrm{F}}^T_{\bm{zb}}) & -\mathfrak{I}(\bm{\mathrm{F}}^T_{\bm{bb}}) & \mathfrak{R}(\bm{\mathrm{F}}_{\bm{bb}})
					\end{array}\right], 
				\end{align}
				where $\bm{\mathrm{F}}_{\bm{xx}}, \bm{\mathrm{F}}_{\bm{yy}}$, and $\bm{\mathrm{F}}_{\bm{zz}}$ are given by
				\begin{align}\label{eq:F_multiple_list_xx_yy_zz}
					\bm{\mathrm{F}}_{\bm{uu}} & =  L (\dot{\bm{A}}_{\bm{u}}^H  \bm{Q}^{-1} \dot{\bm{A}_{\bm{u}}}) \odot (\bm{B}^* \bm{V}^H \bm{R}_X^* \bm{V} \bm{B}) \\
					\nonumber
					& + L (\dot{\bm{A}}_{\bm{u}}^H  \bm{Q}^{-1} \bm{A}) \odot (\bm{B}^* \bm{V}^H \bm{R}_X^* \dot{\bm{V}}_{\bm{u}} \bm{B}) \\
					\nonumber
					& + L (\bm{A}^H  \bm{Q}^{-1} \dot{\bm{A}_{\bm{u}}}) \odot (\bm{B}^* \dot{\bm{V}}_{\bm{u}}^H \bm{R}_X^* \bm{V} \bm{B})\\
					\nonumber
					& + L (\bm{A}^H  \bm{Q}^{-1} \bm{A}) \odot (\bm{B}^* \dot{\bm{V}}_{\bm{u}}^H \bm{R}_X^* \dot{\bm{V}}_{\bm{u}} \bm{B}), \enspace \bm{u} \in \{\bm{x},\bm{y},\bm{z}\}, 
				\end{align}
				$\bm{\mathrm{F}}_{\bm{xy}}, \bm{\mathrm{F}}_{\bm{xz}}$, and $\bm{\mathrm{F}}_{\bm{yz}}$ are given by
				\begin{align}\label{eq:F_multiple_list_xy_xz_yz}
					& \bm{\mathrm{F}}_{\bm{uv}}  =  L (\dot{\bm{A}}_{\bm{u}}^H  \bm{Q}^{-1} \dot{\bm{A}_{\bm{v}}}) \odot (\bm{B}^* \bm{V}^H \bm{R}_X^* \bm{V} \bm{B}) \\
					\nonumber
					& + L (\dot{\bm{A}}_{\bm{u}}^H  \bm{Q}^{-1} \bm{A}) \odot (\bm{B}^* \bm{V}^H \bm{R}_X^* \dot{\bm{V}}_{\bm{v}} \bm{B}) \\
					\nonumber
					& + L (\bm{A}^H  \bm{Q}^{-1} \dot{\bm{A}_{\bm{v}}}) \odot (\bm{B}^* \dot{\bm{V}}_{\bm{u}}^H \bm{R}_X^* \bm{V} \bm{B})\\
					\nonumber
					& + L (\bm{A}^H  \bm{Q}^{-1} \bm{A}) \odot (\bm{B}^* \dot{\bm{V}}_{\bm{u}}^H \bm{R}_X^* \dot{\bm{V}}_{\bm{v}} \bm{B}), \enspace \bm{uv} \in \{\bm{xy},\bm{xz},\bm{yz}\}, 
				\end{align}
				and $\bm{\mathrm{F}}_{\bm{bb}}$, $\bm{\mathrm{F}}_{\bm{xb}}, \bm{\mathrm{F}}_{\bm{yb}}$, and $\bm{\mathrm{F}}_{\bm{zb}}$ are given by
				\begin{align}\label{eq:F_multiple_list_xb_yb_zb}
					\bm{\mathrm{F}}_{\bm{bb}} & = L (\bm{A}^H  \bm{Q}^{-1} \bm{A}) \odot (\bm{V}^H \bm{R}_X^* \bm{V}),\\
					\bm{\mathrm{F}}_{\bm{ub}}  & =  L (\dot{\bm{A}}_{\bm{u}}^H  \bm{Q}^{-1} \bm{A}) \odot (\bm{B}^* \bm{V}^H \bm{R}_X^* \bm{V}) \\
					\nonumber
					& + L (\bm{A}^H  \bm{Q}^{-1} \bm{A}) \odot (\bm{B}^* \dot{\bm{V}}_{\bm{u}}^H \bm{R}_X^* \bm{V}), \enspace \bm{u} \in \{\bm{x},\bm{y},\bm{z}\}, 
				\end{align}
				respectively, with
				\begin{align}\label{eq:VX_partial}
					\dot{\bm{A}_{\bm{u}}} & = \left[\frac{\partial \bm{a}(\bm{l}_1)}{\partial u_1},...,\frac{\partial \bm{a}(\bm{l}_K)}{\partial u_K}\right], \enspace \bm{u} \in \{\bm{x},\bm{y},\bm{z}\},   \\
					\dot{\bm{V}_{\bm{u}}} & = \left[\frac{\partial \bm{v}(\bm{l}_1)}{\partial u_1},...,\frac{\partial \bm{v}(\bm{l}_K)}{\partial u_K}\right], \enspace \bm{u} \in \{\bm{x},\bm{y},\bm{z}\}, 
				\end{align}
				\begin{align}
					\label{eq:AX_partial_element}
					\left(\frac{\partial \bm{a}(\bm{l}_k)}{\partial u_k}\right)_m & = \bm{a}_m(\bm{l}_k)(\frac{u_m^r-u_k}{\|\bm{l}^r_m - \bm{l}_{k}\|^2}+\mathrm{j} \nu \frac{u_m^r-u_k}{\|\bm{l}^r_m - \bm{l}_{k}\|}),   \\
					\label{eq:VX_partial_element}
					\left(\frac{\partial \bm{v}(\bm{l}_k)}{\partial u_k}\right)_n & = \bm{v}_n(\bm{l}_k)(\frac{u_n^t-u_k}{\|\bm{l}^t_n - \bm{l}_{k}\|^2}+\mathrm{j} \nu \frac{u_n^t-u_k}{\|\bm{l}^t_n - \bm{l}_{k}\|}). 
				\end{align}
				In (\ref{eq:AX_partial_element}) and (\ref{eq:VX_partial_element}), $\bm{a}_m(\bm{l}_k)$ and $\bm{v}_n(\bm{l}_k)$ denote the $m$-th element of $\bm{a}(\bm{l}_k)$ and the $n$-th element of $\bm{v}(\bm{l}_k)$, respectively.}
			\begin{proof}
				Please refer to Appendix \ref{Appendix:Proof_Prop_F_deri}.
			\end{proof}
		\end{proposition}

		With Proposition \ref{Pro:F_deri}, we have the complete FIM $\bm{\mathrm{F}}$ related to the interested parameters in $\bm{\theta}$. As a result, the CRB matrix $\bm{\mathrm{C}}$ for estimating $\bm{\theta}$ is given as
		\begin{align}\label{eq:CRB_mat}
			\bm{\mathrm{C}} = \bm{\mathrm{F}}^{-1}.
		\end{align}
		According to (\ref{eq:theta_interested}), we obtain the sum CRB for estimating the position of the $k$-th target $\bm{l}_k$ as
		\begin{align}\label{eq:CRB_position}
			\text{CRB}_{P,k} = \bm{\mathrm{C}}[k,k] + \bm{\mathrm{C}}[k+K,k+K] + \bm{\mathrm{C}}[k+2K,k+2K].
		\end{align}
		
		\begin{remark}\label{remark:path-loss_wo}
			\emph{It is worth noting that the CRB analysis here is more general than that in \cite{khamidullina2021conditional} in the following aspects. First, in (\ref{eq:remain_def_Fisher_Rx}) and (\ref{equ:Rx_data_matrix}) we consider general transmit waveform with sample covariance matrix $\bm{R}_X$ being any arbitrary positive semi-definite matrix, while \cite{khamidullina2021conditional} focused on the special orthogonal waveforms with $\bm{R}_X = \bm{I}$. Next, in (\ref{equ:Rx_data_matrix}) we consider unknown noise and interference covariance matrix $\bm{Q}$, while \cite{khamidullina2021conditional} assumed WGN model with $\bm{Q} = \sigma^2  \bm{I}$, where $\sigma^2$ is the noise power. Furthermore,  we consider the general varying channel amplitude across antennas in (\ref{equ:steer_Rx}) and (\ref{equ:steer_Tx}), while \cite{khamidullina2021conditional} considered a simplified constant-amplitude model with $A_1^r(\bm{l}_k)=...=A_M^r(\bm{l}_k) \triangleq A^r(\bm{l}_k)$ and $A_1^t(\bm{l}_k)=...=A_N^t(\bm{l}_k) \triangleq A^t(\bm{l}_k)$, where the round-trip path loss effects are included in the target complex reflection coefficients.}
	\end{remark}

		\section{Near-Field Localization via 3D-ACO} \label{Section_estimators}
		This section presents a practical near-field localization algorithm via 3D-ACO based on (\ref{equ:data_compact_rx}) following the ML criterion.
		
		 Recall that  $\bm{y} \triangleq \operatorname{vec}(\bm{Y}) \sim \mathcal{CN}(\bm{\mu},\bm{C})$, with $\bm{\mu}$ and $\bm{C}$ given in (\ref{equ:Rx_data_mat_DGC}) and (\ref{equ:cov_C}), respectively. Here, we omit $\tilde{\bm{\theta}}$ for simplicity. Thus, the negative log-likelihood function of $\bm{y}$ is 
		\begin{align}\label{eq:initial_neg_log_like}
			- \ln f(\bm{y}) = -\ln \frac{1}{\pi^{L M} |\bm{C}|} + (\bm{y}-\bm{\mu})^H \bm{C}^{-1} (\bm{y}-\bm{\mu}).
		\end{align}
		With the constant term omitted, the negative log-likelihood in (\ref{eq:initial_neg_log_like}) is re-expressed as
		\begin{align}\label{eq:neg_log_like}
			\nonumber
			f_1(\bm{Q},\{b_k\},\{\bm{l}_k\}) & = L \ln |\bm{Q}| + \operatorname{tr}\left[(\bm{A} \bm{B} \bm{V}^T \bm{X}) \right. \\
			& \left. (\bm{Y} - \bm{A} \bm{B} \bm{V}^T \bm{X})^H \bm{Q}^{-1}\right].
		\end{align}
		Let $\bm{W} \triangleq (\bm{Y} - \bm{A} \bm{B} \bm{V}^T \bm{X})(\bm{Y} - \bm{A} \bm{B} \bm{V}^T \bm{X})^H$.
		We then have
		\begin{align}\label{eq:f_1_expression}
			f_1(\bm{Q},\{b_k\},\{\bm{l}_k\}) = L \ln |\bm{Q}| + \operatorname{tr}(\bm{W} \bm{Q}^{-1}).
		\end{align}
		By maximizing the log-likelihood or equivalently minimizing $f_1$ with respect to $\bm{Q}$, we obtain the estimate of $\bm{Q}$ as 
		\begin{align}\label{eq:Est_Q_opt}
			\bm{Q}^\star = \frac{1}{L} \bm{W}.
		\end{align}
		Substituting it back to (\ref{eq:f_1_expression}), we have
		\begin{align}\label{eq:f_1_with_W}
			f_1 (\frac{1}{L} \bm{W}, \{b_k\}, \{\bm{l}_k\}) = L \ln (\frac{1}{L})^M + L \ln |\bm{W}| + L M.
		\end{align}
		By ignoring the first and the third constant terms, the minimization of $f_1$ in (\ref{eq:f_1_with_W}) is equivalent to minimizing
		\begin{align}\label{equ:f2_AML}
			& f_2(\{b_k\}, \{\bm{l}_k\})   = L \ln |\bm{W}| \\
			\nonumber
			& = L \ln |(\bm{Y} - \bm{A} \operatorname{diag}(\bm{b}) \bm{V}^T \bm{X})(\bm{Y} - \bm{A} \operatorname{diag}(\bm{b}) \bm{V}^T \bm{X})^H|.
		\end{align}
		
		Next, we seek to minimize $f_2$ with respect to 3D target locations $\{\bm{l}_k\}$ and target reflection coefficients $\bm{b}$. 
		We first obtain the estimation of the complex reflection coefficient vector $\bm{b}$ with given $\{\bm{l}_k\}$. Notice that the received data matrix in (\ref{equ:data_compact_rx}) is in canonical form known as the diagonal growth curve (DGC) model \cite{xu2006diagonal}. Therefore, we apply the approximate maximum likelihood (AML) estimator in \cite{xu2006diagonal} to obtain an AML estimate of $\bm{b}$ as
		\begin{align}\label{equ:AML_Est_b}
			\bm{b}  =  \left[(\bm{A}^H \bm{\mathrm{J}}^{-1} \bm{A}) \odot (\bm{\mathrm{S}} \bm{\mathrm{S}}^H)^T \right]^{-1} \operatorname{vecd}\left(\bm{A}^H \bm{\mathrm{J}}^{-1} \bm{Y} \bm{\mathrm{S}}^H\right), 
		\end{align}
		where 
		$\operatorname{vecd}(\cdot)$ denotes a column vector formed by the diagonal elements of a given matrix, and 
		\begin{align}
			\label{equ:def_S}
			\bm{\mathrm{S}} & = \bm{V}^T \bm{X}, \\
			\label{equ:def_J}
			\bm{\mathrm{J}} & = \frac{1}{L} \bm{Y} \bm{Y}^H - \frac{1}{L} \bm{Y} \bm{\mathrm{S}}^H (\bm{\mathrm{S}} \bm{\mathrm{S}}^H)^{-1} \bm{\mathrm{S}} \bm{Y}^H.
		\end{align}
		Combining 
		(\ref{equ:f2_AML})-(\ref{equ:def_J}), 
		we obtain the concentrated negative log-likelihood function $f_3(\{\bm{l}_k\})$ as
		\begin{align}\label{equ:f1_neg_log}
			f_3(\{\bm{l}_k\}) = L \ln |(\bm{Y} - \bm{A} \operatorname{diag}(\bm{b}) \bm{\mathrm{S}})(\bm{Y} - \bm{A} \operatorname{diag}(\bm{b}) \bm{\mathrm{S}})^H|,
		\end{align}
		with $\bm{b}$, $\bm{\mathrm{S}}$, and $\bm{\mathrm{J}}$ given in (\ref{equ:AML_Est_b}), (\ref{equ:def_S}) and (\ref{equ:def_J}), respectively.
		In (\ref{equ:f1_neg_log}), the number of unknowns has been reduced to $3 K$. 
		
		We now proceed to resolve the locations of all the targets, i.e., $\{\bm{l}_k\}$ by minimizing $f_3(\{\bm{l}_k\})$. Towards this end, we apply the cyclic optimization technique \cite{li2008mimo}, which is summarized in Algorithm 1 and is implemented in a cyclic manner. To start with, we first estimate the location of a single target by finding the minimizer of $f_3$ in (\ref{equ:f1_neg_log}) via a 3D exhaustive search (Steps 1-3 in Algorithm 1). Then, if $K_{\text{max}}>1$, we gradually increase the expected number of targets $\hat{K}$. Under given $\hat{K}$, we estimate the locations of the $\hat{K}$ targets cyclically. In other words, in each iteration, we estimate the location of each target as the one minimizing $f_3$ in (\ref{equ:f1_neg_log}) via a 3D exhaustive search, in which the estimated locations of the other targets are considered fixed (Steps 6-19 in Algorithm 1). The cyclic iteration terminates until some convergence criterion is met, e.g., the relative change of $f_3$ between two consecutive iterations is less than some pre-determined threshold $\epsilon$ (Step 11 in Algorithm 1). Finally, the outer iteration terminates when the number of expected targets $\hat{K}$ reaches a prescribed maximum number $K_{\text{max}}$ (Step 4 in Algorithm 1).
		
		\begin{algorithm}[t]
			\label{alg:3D_CO}
			\textbf{Algorithm 1} 3D Cyclic Optimization for targets localization
			\hrule
			\begin{itemize}
				\item \textbf{Input}: $\bm{X}$, $\bm{Y}$, $K_{\text{max}}$, $\epsilon$, and $\bm{L} \in \mathbb{R}^{3 \times K_{\text{max}}}$
				
				\begin{algorithmic}[1]
					\STATE $\hat{K} \gets 1$ 
					\STATE Estimate $\bm{l}_1 = \left[x_1,y_1,z_1\right]^T$ as the one minimizing $f_3$ in (\ref{equ:f1_neg_log})
					\STATE $\bm{L}\left[:,1\right] \gets \bm{l}_1$ 
					\WHILE  {$\hat{K} < K_{\text{max}}$}
					\STATE $\hat{K} \gets \hat{K}+1$
					\STATE Estimate $\bm{l}_{\hat{K}} = \left[x_{\hat{K}},y_{\hat{K}},z_{\hat{K}}\right]^T$ as the one minimizing $f_3$ in (\ref{equ:f1_neg_log}) with fixed $\{\bm{l}_m\}_{m=1}^{\hat{K}-1}$ 
					\STATE Update $f_3$ based on  $\{\bm{l}_m\}_{m=1}^{\hat{K}}$ with newly estimated $\bm{l}_{\hat{K}}$
					\STATE $\bm{L}\left[:,\hat{K}\right] \gets \bm{l}_{\hat{K}}$ 
					\STATE $f_\text{old} \gets  f_3(\{\bm{l}_m\}_{m=1}^{\hat{K}}) + 2\epsilon, \text{ } f_\text{new} \gets f_3(\{\bm{l}_m\}_{m=1}^{\hat{K}})$
					\STATE $  p \gets 1$
					\WHILE {$\left(f_\text{old} - f_\text{new} > \epsilon\right)$} 
				\STATE $ f_\text{old} \gets f_\text{new}$
				\STATE Estimate $\bm{l}_p = \left[x_p,y_p,z_p\right]^T$ as the one minimizing $f_3$ in (\ref{equ:f1_neg_log}) with fixed $\{\bm{l}_m\}_{m=1,m \neq p}^{\hat{K}}$ 
				\STATE Update $f_3$ based on $\{\bm{l}_m\}_{m=1}^{\hat{K}}$ with newly estimated $\bm{l}_p$
				\STATE $\bm{L}\left[:,p\right] \gets \bm{l}_p$ 
				\STATE $f_\text{new} \gets f_3(\{\bm{l}_m\}_{m=1}^{\hat{K}})$
				\STATE $p \gets \mod (p,\hat{K})+1$
				\ENDWHILE
				\ENDWHILE
				\STATE $\bm{l}_k^{\text{ACO}} = \bm{L}\left[:,k\right], k = 1,2,..., K_\text{max}$
			\end{algorithmic}
			\item \textbf{Output}: $\{\bm{l}_k^{\text{ACO}}\}_{k=1}^{K_{\text{max}}}$
		\end{itemize}
	\end{algorithm}

		%
		%
		%
		%
		%
		%
	
	The complexity of each 3D exhaustive search is typically large. To reduce the complexity, we first specify the initial search grid and obtain a coarse estimate of the location of the target. Then, based on the obtained coarse estimation, we iteratively find better estimation around the previous estimate with reduced grid search size in each iteration.
	
	\begin{remark}
		\emph{Note that in the estimators design, we assume that the number of the targets is known as $K$ in advance and accordingly set it as the maximum searching number $K_{\text{max}}$ in Algorithm 1. The estimator can be extended to the case when $K$ is not  \textit{a-priori} known by combining it with the model order selection rules \cite{stoica2004model}. }
	\end{remark}

	\section{Numerical Results} \label{Section_results}
	
	This section provides numerical results to validate the derived near-field CRB and evaluate the performance of the proposed estimator 3D-ACO for 3D targets localization as compared with the derived CRB.

	\subsection{CRB Behavior Analysis}
	We first demonstrate the necessity of considering distance-dependent channel amplitude variation in (\ref{equ:steer_Rx}) and (\ref{equ:steer_Tx}).
	For comparison, we consider the conventional analysis without varying amplitude across antennas \cite{khamidullina2021conditional}. Towards this end, 
	to ensure fair comparison, we normalize the constant-amplitude term, i.e., $A^r(\bm{l}_k)$ and $A^t(\bm{l}_k)$, into the complex reflection coefficient $\{b_k\}$ and accordingly get the complex reflection coefficient with round-trip path-loss, denoted as $\{\tilde{b}_k\}$, as discussed in Remark \ref{remark:path-loss_wo}. 
	Specifically, the CRB without considering distance-dependent channel amplitude variations is derived based on the following received signal model:
	\begin{align}\label{equ:Rx_data_matrix_wo_pl}
		\tilde{\bm{Y}} = \sum_{k = 1}^{K} \tilde{\bm{a}}(\bm{l}_k) \tilde{b}_k \tilde{\bm{v}}^T(\bm{l}_k) \bm{X} + \bm{Z},
	\end{align}
	where $\left[\tilde{\bm{a}}(\bm{l}_k)\right]_m = e^{-\mathrm{j} \nu \|\bm{l}^r_m - \bm{l}_{k}\|}, \left[\tilde{\bm{v}}(\bm{l}_k)\right]_n = e^{-\mathrm{j} \nu \|\bm{l}^t_n - \bm{l}_{k}\|}$, and 
	\begin{align}\label{eq:tilde_b_k}
		\tilde{b}_k = \left(\frac{\lambda}{4 \pi}\right)^2   \frac{1}{\|\bm{l}^r_o-\bm{l}_{k}\|} \frac{1}{\|\bm{l}^t_o-\bm{l}_{k}\|} b_k, 
	\end{align}
	where $\bm{l}^r_o$ and $\bm{l}^t_o$ are the 3D coordinates of the reference point at Rx and Tx, respectively.
	Here, we choose the path-loss with respect to the reference points and accordingly obtain the complex reflection coefficient with round-trip path-loss $\{\tilde{b}_k\}$.
	\begin{figure}[t]
		\centering
		\includegraphics[width=2.2in]{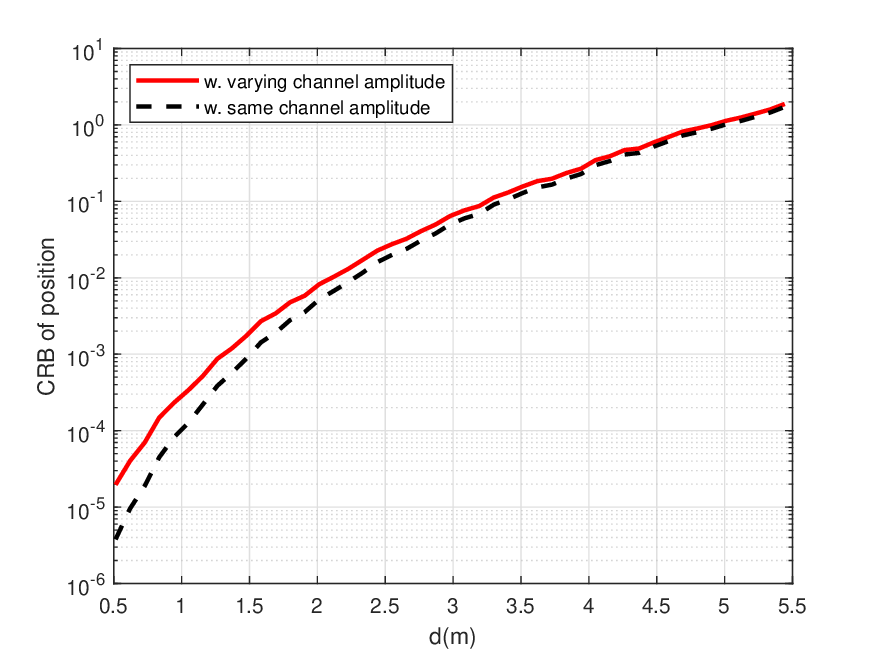}
		\centering
		\caption{The comparison of the CRB of 3D target positions in (\ref{eq:CRB_position}) under our considered setup versus that with same channel amplitudes across antennas.}
		\label{fig:CRB_wo_pl_analysis}
	\end{figure}
	
	Considering a mono-static MIMO radar setup, which is a special case of Fig. \ref{fig:sm}(a), in which both Rx and Tx are completely overlapped with each other. Accordingly, we assume $\bm{l}^r_o = \bm{l}^t_o = \left[0,0,0\right]^T$.
	Without loss of generality, we further consider a UPA in a rectangular shape with $M = N = 16 \times 768 = 12288$. The two sides of the UPA are in parallel with $x$-axis and $y$-axis, respectively, and the center of the UPA is the original point of the $x-y$ plane. The spacing between adjacent antennas is half-wavelength with $f_c = 28$ GHz. We further consider the simplest WGN model, i.e., $\bm{Q} = \sigma^2 \bm{I}$ with $\sigma^2 = 0.001$, and adopt an isotropic transmission with unit power, i.e., $\bm{R}_X \sim \bm{I}$ with $L = 256$. Assume that there exists a single point target lying on the $z$-axis with coordinate $(0,0,d)$.

As we can see in Fig. \ref{fig:CRB_wo_pl_analysis}, when the target is close to the UPA, the CRB of position estimation without considering distance-dependent channel amplitude significantly deviates from our derived CRB by taking it into account.
This justifies the necessity of considering distance-dependent channel amplitude in channel modeling. We also observe that when the target is moving away from the UPA, two different models will converge. This shows that the simplified model without considering distance-dependent channel amplitude is valid as long as the target is distant enough from the transceiver. 


\subsection{3D-ACO Evaluation with the CRB}
We proceed to evaluate the performance of the proposed estimator 3D-ACO with the derived CRB. Throughout the whole subsection, the SNR is defined as
$\text{SNR} = \frac{\mathbb{E}\left[ \|\bm{A}\bm{B}\bm{V}^T\bm{x}_l\|^2 \right]}{\mathbb{E} \left[\|\bm{z}_l\|^2\right]},$
which can be approximated via
$\tilde{\text{SNR}} = \frac{ \sum_{l=1}^{L} \|\bm{A}\bm{B}\bm{V}^T\bm{x}_l\|^2 }{\sum_{l=1}^{L}\|\bm{z}_l\|^2},$
as long as $L$ is sufficiently long.

\begin{figure}[t]
\centering
\includegraphics[width=2.2in]{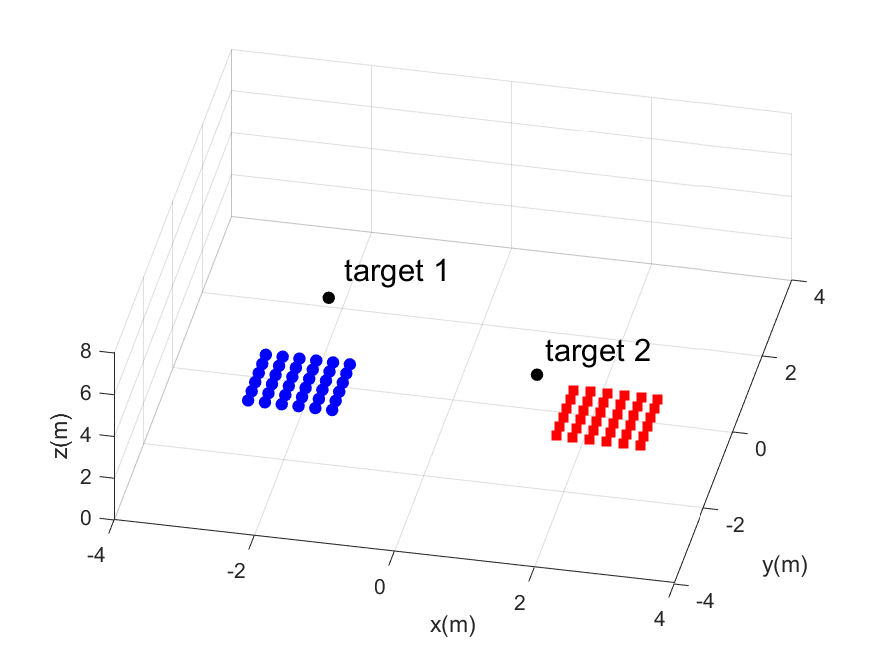}
\centering
\caption{Configuration of the antenna array and the relative location of the targets when there exists two targets. The red square array denotes the Tx and the blue circle array denotes the Rx.}
\label{fig:two_target_scen}
\end{figure}
Fig. \ref{fig:two_target_scen} shows the antenna configuration. 
The blue UPA corresponds to the Rx array while the red UPA corresponds to the Tx array. The UPA is square with $M = N = 6^2 = 36$, and the spacing between adjacent antennas is half-wavelength with $f_c = 0.625$ GHz.  
We adopt the WGN model and isotropic transmission for simplicity, i.e., $\bm{Q} = \sigma^2 \bm{I}$ and $\bm{R}_X \sim \bm{I}$ with $L = 52$, respectively. Besides, the pre-determined threshold $\epsilon$ is set to be $10^{-5}$ in Algorithm 1.

\begin{figure}[t]
\centering
\setlength{\abovecaptionskip}{+4mm}
\setlength{\belowcaptionskip}{+1mm}
\subfigure[MSE versus SNR for target 1.]{ \label{fig:MSE_1st}
\includegraphics[width=2.1in]{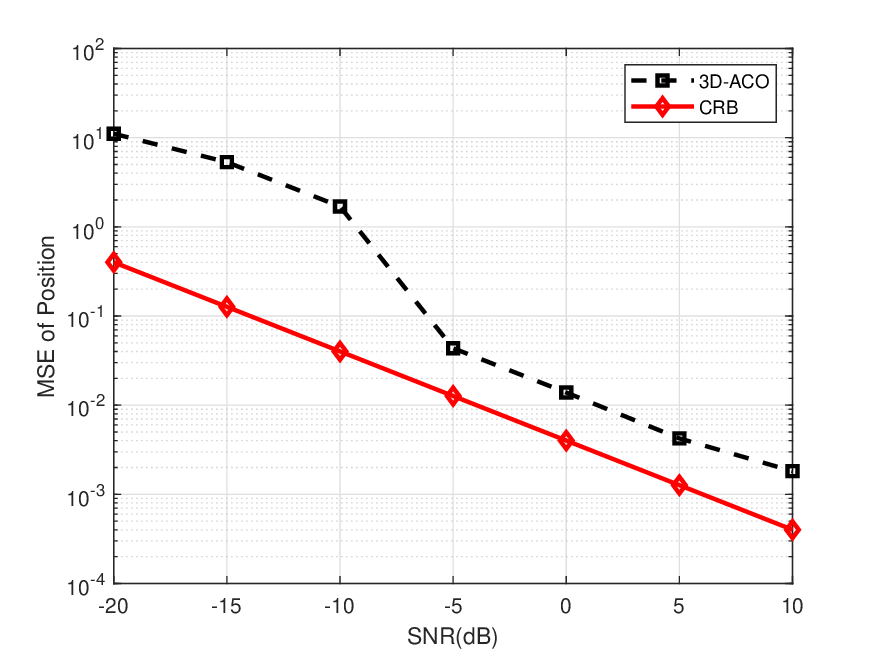}}
\subfigure[MSE versus SNR for target 2.]{ \label{fig:MSE_2rd}
\includegraphics[width=2.1in]{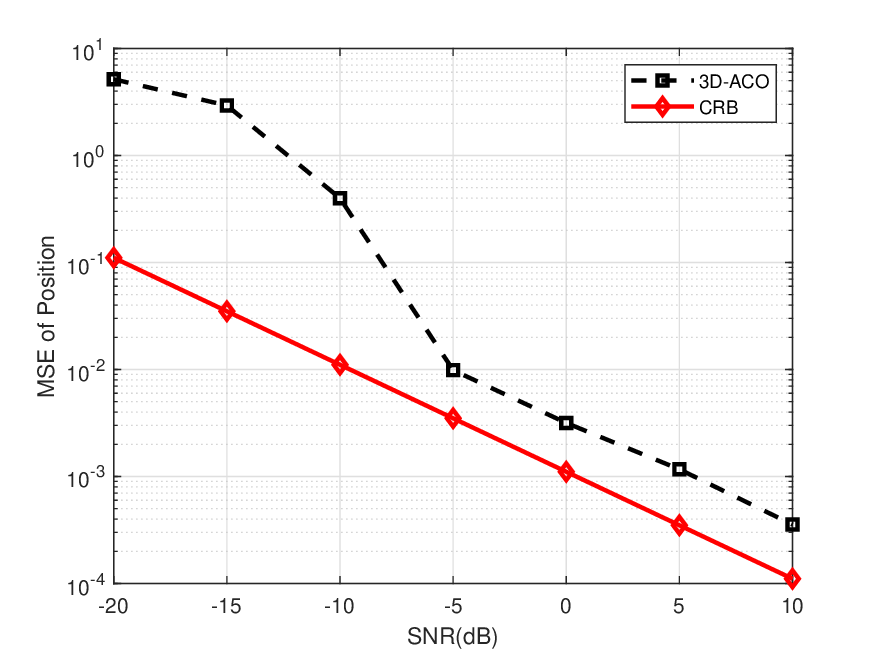}}
\caption{The performance of the proposed 3D-ACO estimator versus the derived CRB in terms of MSE with respect to SNR with $L = 52, M = N = 36$, which are obtained by averaging over 100 Monte Carol realizations. }
\label{fig:multiple_nf_MSE}
\end{figure}

Fig. \ref{fig:multiple_nf_MSE} shows the MSE of the target positions versus the SNR for the two targets. It is observed that when the SNR becomes large, both the CRB of the positions and the performance of 3D-ACO estimator for two different targets become better, leading to smaller MSE. It is also observed that the CRB for two targets are different in terms of their values, as the CRB for each target is dependent not only on the antenna configuration, but also on the relative position of all the targets needed to be localized. Finally, it is observed that in the high SNR regime, the performance of the 3D-ACO estimator is close to the derived CRB, validating its superior performance in multi-target localization.

\section{Conclusion}
This paper studied a near-field MIMO radar system for multi-target localization by considering the exact spherical wavefront model with both channel phase and amplitude variations across antennas. Under this setup, we derived the CRB for estimating the 3D target locations, which is applicable in the scenario with general transmit waveforms and unknown noise and interference covariance matrix. Next, we developed a practical 3D-ACO localization algorithm based on the ML criterion. Numerical results showed that the consideration of antenna-specific channel amplitude variation is crucial for the CRB performance, particularly when the target is close to the transceivers. It was also shown that the proposed 3D-ACO algorithm achieved MSE close to the CRB. The CRB analysis in this paper is expected to have wide applications to facilitate the adaptive transmit design in near-field MIMO sensing and MIMO ISAC for future research.

\appendix

\subsection{Proof of Proposition \ref{Pro:overall_fisher}}\label{Appendix:Proof_Prop_1}

The received vectorized data matrix $\bm{y} \sim \mathcal{CN}(\bm{\mu}(\tilde{\bm{\theta}}),\bm{C}(\tilde{\bm{\theta}}))$, with $\bm{\mu}(\tilde{\bm{\theta}})$ and $\bm{C}(\tilde{\bm{\theta}})$ given in (\ref{equ:Rx_data_mat_DGC}) and (\ref{equ:cov_C}), respectively. Then, according to Lemma \ref{lemma:Fisher_deri}, we have
\begin{align}\label{equ:Fisher_tilde_F_deri}
	\nonumber
	\tilde{\bm{\mathrm{F}}}(\tilde{\bm{\theta}}_i,\tilde{\bm{\theta}}_j)  & = \operatorname{tr}\left[\bm{C}^{-1}(\tilde{\bm{\theta}}) \frac{\partial \bm{C}(\tilde{\bm{\theta}})}{\partial \tilde{\bm{\theta}}_i} \bm{C}^{-1}(\tilde{\bm{\theta}}) \frac{\partial \bm{C}(\tilde{\bm{\theta}})}{\partial \tilde{\bm{\theta}}_j}\right]   \\
	& + 2 \mathfrak{R}\left[  (\frac{\partial \bm{\mu}}{\partial \tilde{\bm{\theta}}_i})^H \bm{C}^{-1}(\tilde{\bm{\theta}}) \frac{\partial \bm{\mu}}{\partial \tilde{\bm{\theta}}_j}\right].
\end{align}
As $\bm{C}(\tilde{\bm{\theta}})$ is a block diagonal matrix, the inverse of $\bm{C}(\tilde{\bm{\theta}})$ can be obtained based on the inverse of each individual block matrix $\bm{Q}$ in $\bm{C}(\tilde{\bm{\theta}})$. As a result, the first term in (\ref{equ:Fisher_tilde_F_deri}) can be rewritten as
\begin{align}
	\nonumber
	\operatorname{tr}\left\{ \left[\begin{array}{ccc}
		\bm{Q}^{-1} \frac{\partial \bm{Q}}{\partial \tilde{\bm{\theta}}_i} \bm{Q}^{-1} \frac{\partial \bm{Q}}{\partial \tilde{\bm{\theta}}_j}  & \bm{0} & \bm{0} \\
		\bm{0}&  \ddots & \bm{0}\\
		\bm{0}  &\bm{0} & \bm{Q}^{-1} \frac{\partial \bm{Q}}{\partial \tilde{\bm{\theta}}_i} \bm{Q}^{-1} \frac{\partial \bm{Q}}{\partial \tilde{\bm{\theta}}_j} 
	\end{array}\right]  \right\},
\end{align}
which corresponds to the first term $L \operatorname{tr}\left[\bm{Q}^{-1} \frac{\partial \bm{Q}}{\partial \tilde{\bm{\theta}}_i} \bm{Q}^{-1} \frac{\partial \bm{Q}}{\partial \tilde{\bm{\theta}}_j}\right]$ in (\ref{eq:overall_fisher}).
The second term in (\ref{equ:Fisher_tilde_F_deri}) is rewritten as
\begin{align}\label{equ:Gauss_CRB_sim}
\nonumber
2 \mathfrak{R} & \left[(\frac{\partial (\bm{A} \bm{B} \bm{V}^T \bm{x}_1)}{\partial \tilde{\bm{\theta}}_i})^H,...,(\frac{\partial (\bm{A} \bm{B} \bm{V}^T \bm{x}_L)}{\partial \tilde{\bm{\theta}}_i})^H\right] \\ 
\nonumber
& \left[\begin{array}{ccc}
\bm{Q}^{-1} & \bm{0} &  \bm{0} \\
\bm{0} & \ddots &  \bm{0}\\
\bm{0} & \bm{0} & \bm{Q}^{-1}  
\end{array}\right]  \left[\begin{array}{c}
\frac{\partial (\bm{A} \bm{B} \bm{V}^T \bm{x}_1)}{\partial \tilde{\bm{\theta}}_j}  \\
\vdots \\
\frac{\partial (\bm{A} \bm{B} \bm{V}^T \bm{x}_L)}{\partial \tilde{\bm{\theta}}_j}
\end{array}\right]   \\
\nonumber
\end{align}
\begin{align}
	\nonumber
= & 2 \mathfrak{R} \left\{ \sum_{l=1}^{L} (\frac{\partial (\bm{A} \bm{B} \bm{V}^T \bm{x}_l)}{\partial \tilde{\bm{\theta}}_i})^H \bm{Q}^{-1} (\frac{\partial (\bm{A} \bm{B} \bm{V}^T \bm{x}_l)}{\partial \tilde{\bm{\theta}}_j}) \right\} \\
= & 2 \mathfrak{R} \operatorname{tr} \left[(\frac{\partial (\bm{A} \bm{B} \bm{V}^T \bm{X})}{\partial \tilde{\bm{\theta}}_i})^H \bm{Q}^{-1} (\frac{\partial (\bm{A} \bm{B} \bm{V}^T \bm{X})}{\partial \tilde{\bm{\theta}}_j}) \right],
\end{align}
which corresponds to the second term in (\ref{eq:overall_fisher}). This thus completes the proof.

\subsection{Proof of Proposition \ref{Pro:F_deri}}\label{Appendix:Proof_Prop_F_deri}
As $\bm{\theta}$ in (\ref{eq:theta_interested}) does not contain any unknowns in $\bm{Q}$, the first term in (\ref{eq:overall_fisher}) vanishes. Thus, we have
\begin{align}\label{eq:interested_fisher}
\bm{\mathrm{F}}  (\bm{\theta}_i,\bm{\theta}_j)   = 2 \mathfrak{R} \operatorname{tr} \left[ (\frac{\partial (\bm{A} \bm{B} \bm{V}^T \bm{X})}{\partial \bm{\theta}_i})^H \bm{Q}^{-1} (\frac{\partial (\bm{A} \bm{B} \bm{V}^T \bm{X})}{\partial \bm{\theta}_j}) \right].
\end{align} 

The overall structure of the FIM is given below, in which 25 blocks exist in total. 
\begin{align}
\nonumber
\left[\begin{array}{rrrrr}
\bm{\mathrm{F}}(\bm{x},\bm{x}) & \bm{\mathrm{F}}(\bm{x},\bm{y})& \bm{\mathrm{F}}(\bm{x},\bm{z})& \bm{\mathrm{F}}(\bm{x},\bm{b}_\text{R})&\bm{\mathrm{F}}(\bm{x},\bm{b}_\text{I})\\
\bm{\mathrm{F}}(\bm{y},\bm{x}) & \bm{\mathrm{F}}(\bm{y},\bm{y})& \bm{\mathrm{F}}(\bm{y},\bm{z})& \bm{\mathrm{F}}(\bm{y},\bm{b}_\text{R})&\bm{\mathrm{F}}(\bm{y},\bm{b}_\text{I})\\
\bm{\mathrm{F}}(\bm{z},\bm{x}) & \bm{\mathrm{F}}(\bm{z},\bm{y})& \bm{\mathrm{F}}(\bm{z},\bm{z})& \bm{\mathrm{F}}(\bm{z},\bm{b}_\text{R})&\bm{\mathrm{F}}(\bm{z},\bm{b}_\text{I}) \\
\bm{\mathrm{F}}(\bm{b}_\text{R},\bm{x}) & \bm{\mathrm{F}}(\bm{b}_\text{R},\bm{y})& \bm{\mathrm{F}}(\bm{b}_\text{R},\bm{z})& \bm{\mathrm{F}}(\bm{b}_\text{R},\bm{b}_\text{R})&\bm{\mathrm{F}}(\bm{b}_\text{R},\bm{b}_\text{I})\\
\bm{\mathrm{F}}(\bm{b}_\text{I},\bm{x}) & \bm{\mathrm{F}}(\bm{b}_\text{I},\bm{y})& \bm{\mathrm{F}}(\bm{b}_\text{I},\bm{z})& \bm{\mathrm{F}}(\bm{b}_\text{I},\bm{b}_\text{R})&\bm{\mathrm{F}}(\bm{b}_\text{I},\bm{b}_\text{I})\\
\end{array}\right]
\end{align}
However, the FIM is always symmetric positive semidefinite, and therefore, we can reduce the number of blocks needed to check.

First, we find $\bm{\mathrm{F}}(\bm{x},\bm{x})$. Towards this end, we have 
\begin{align}\label{equ:Gauss_CRB_x}
\bm{\mathrm{F}}(x_i,x_j) = 2 \mathfrak{R} \operatorname{tr} \left[ (\frac{\partial (\bm{A} \bm{B} \bm{V}^T \bm{X})}{\partial x_i})^H \bm{Q}^{-1} (\frac{\partial (\bm{A} \bm{B} \bm{V}^T \bm{X})}{\partial x_j}) \right].
\end{align}
Notice that for arbitrary matrix $\bm{A}$ and $\bm{B}$ depending on parameter $\rho$, we have
\begin{align} \label{equ:mat_der}
\frac{\partial (\bm{A}\bm{B})}{\partial \rho} = \frac{\partial \bm{A}}{\partial \rho} \bm{B} + \bm{A} \frac{\partial \bm{B}}{\partial \rho}.
\end{align}
Thus, it follows that
\begin{align}\label{equ:Gauss_CRB_x_left}
\frac{\partial\left(\bm{A B} \bm{V}^T \boldsymbol{X}\right)}{\partial x_i}=\dot{\bm{A}_{\bm{x}}} \bm{e}_i \bm{e}_i^T \bm{B} \bm{V}^T \boldsymbol{X}+\bm{A B} \bm{e}_i \bm{e}_i^T \dot{\bm{V}}_{\bm{x}}^T \boldsymbol{X},
\end{align}
where $\bm{e}_i$ denotes the $i$-th column of the identity matrix. Then 
\begin{align}\label{equ:Gauss_CRB_x1}
\nonumber
\bm{\mathrm{F}}(x_i,&x_j)   = 2 \mathfrak{R} \operatorname{tr} \left[ \left(\dot{\bm{A}_{\bm{x}}} \bm{e}_i \bm{e}_i^T \bm{B} \bm{V}^T \boldsymbol{X}+\bm{A B} \bm{e}_i \bm{e}_i^T \dot{\bm{V}}_{\bm{x}}^T \boldsymbol{X}\right)^H \right.\\
& \left. \bm{Q}^{-1} \left(\dot{\bm{A}_{\bm{x}}} \bm{e}_j \bm{e}_j^T \bm{B} \bm{V}^T \boldsymbol{X}+\bm{A B} \bm{e}_j \bm{e}_j^T \dot{\bm{V}}_{\bm{x}}^T \boldsymbol{X}\right) \right].
\end{align}
Notice that 
\begin{align}
\nonumber
\operatorname{tr} & \left[ \left(\dot{\bm{A}_{\bm{x}}} \bm{e}_i \bm{e}_i^T \bm{B} \bm{V}^T \boldsymbol{X}\right)^H \bm{Q}^{-1} \left(\dot{\bm{A}_{\bm{x}}} \bm{e}_j \bm{e}_j^T \bm{B} \bm{V}^T \boldsymbol{X}\right) \right] \\
\nonumber
& = \bm{e}_i^T (\dot{\bm{A}}_{\bm{x}}^H  \bm{Q}^{-1} \dot{\bm{A}_{\bm{x}}}) \bm{e}_j \bm{e}_j^T (\bm{B} \bm{V}^T \bm{X} \bm{X}^H \bm{V}^* \bm{B}^H) \bm{e}_i \\
\nonumber
& = L (\dot{\bm{A}}_{\bm{x}}^H  \bm{Q}^{-1} \dot{\bm{A}_{\bm{x}}})_{ij} (\bm{B}^* \bm{V}^H \bm{R}_X^* \bm{V} \bm{B})_{ij},
\end{align}
which corresponds to the first term of the cross product in (\ref{equ:Gauss_CRB_x1}). Here, $\bm{M}_{ij}$ denotes the $(i,j)$-th element of $\bm{M}$. The other three terms have similar forms. Thus, 
$\bm{\mathrm{F}}(\bm{x},\bm{x}) = 2 \mathfrak{R}(\bm{\mathrm{F}}_{\bm{xx}})$, where
\begin{align}\label{equ:Gauss_CRB_Fxx}
\nonumber
\bm{\mathrm{F}}_{\bm{xx}} & = L (\dot{\bm{A}}_{\bm{x}}^H  \bm{Q}^{-1} \dot{\bm{A}_{\bm{x}}}) \odot (\bm{B}^* \bm{V}^H \bm{R}_X^* \bm{V} \bm{B}) \\
\nonumber
& + L (\dot{\bm{A}}_{\bm{x}}^H  \bm{Q}^{-1} \bm{A}) \odot (\bm{B}^* \bm{V}^H \bm{R}_X^* \dot{\bm{V}}_{\bm{x}} \bm{B}) \\
\nonumber
& + L (\bm{A}^H  \bm{Q}^{-1} \dot{\bm{A}_{\bm{x}}}) \odot (\bm{B}^* \dot{\bm{V}}_{\bm{x}}^H \bm{R}_X^* \bm{V} \bm{B})\\
& + L (\bm{A}^H  \bm{Q}^{-1} \bm{A}) \odot (\bm{B}^* \dot{\bm{V}}_{\bm{x}}^H \bm{R}_X^* \dot{\bm{V}}_{\bm{x}} \bm{B}).
\end{align}
Accordingly, $\bm{\mathrm{F}}_{\bm{yy}}$ and $\bm{\mathrm{F}}_{\bm{zz}}$ can be found by changing $\bm{x}$ in (\ref{equ:Gauss_CRB_Fxx}) into $\bm{y}$ and $\bm{z}$, respectively. Accordingly, $\bm{\mathrm{F}}(\bm{y},\bm{y}) = 2 \mathfrak{R}(\bm{\mathrm{F}}_{\bm{yy}})$, $\bm{\mathrm{F}}(\bm{z},\bm{z}) = 2 \mathfrak{R}(\bm{\mathrm{F}}_{\bm{zz}})$.

Second, we find $\bm{\mathrm{F}}(\bm{x},\bm{y})$. In this regard, one can similarly derive that
\begin{align}\label{equ:Gauss_CRB_xy}
\nonumber
\bm{\mathrm{F}}(x_i,&y_j)
= 2 \mathfrak{R} \operatorname{tr} \left[ \left(\dot{\bm{A}_{\bm{x}}} \bm{e}_i \bm{e}_i^T \bm{B} \bm{V}^T \boldsymbol{X}+\bm{A B} \bm{e}_i \bm{e}_i^T \dot{\bm{V}}_{\bm{x}}^T \boldsymbol{X}\right)^H \right.\\
& \left. \bm{Q}^{-1} \left(\dot{\bm{A}_{\bm{y}}} \bm{e}_j \bm{e}_j^T \bm{B} \bm{V}^T \boldsymbol{X}+\bm{A B} \bm{e}_j \bm{e}_j^T \dot{\bm{V}}_{\bm{y}}^T \boldsymbol{X}\right) \right],
\end{align}
and find that $\bm{\mathrm{F}}(\bm{x},\bm{y}) = 2 \mathfrak{R}(\bm{\mathrm{F}}_{\bm{xy}})$, where
\begin{align}\label{equ:Gauss_CRB_XY}
\nonumber
\bm{\mathrm{F}}_{\bm{xy}} &= L (\dot{\bm{A}}_{\bm{x}}^H  \bm{Q}^{-1} \dot{\bm{A}_{\bm{y}}}) \odot (\bm{B}^* \bm{V}^H \bm{R}_X^* \bm{V} \bm{B}) \\
\nonumber
& + L (\dot{\bm{A}}_{\bm{x}}^H  \bm{Q}^{-1} \bm{A}) \odot (\bm{B}^* \bm{V}^H \bm{R}_X^* \dot{\bm{V}}_{\bm{y}} \bm{B}) \\
\nonumber
& + L (\bm{A}^H  \bm{Q}^{-1} \dot{\bm{A}_{\bm{y}}}) \odot (\bm{B}^* \dot{\bm{V}}_{\bm{x}}^H \bm{R}_X^* \bm{V} \bm{B})\\
& + L (\bm{A}^H  \bm{Q}^{-1} \bm{A}) \odot (\bm{B}^* \dot{\bm{V}}_{\bm{x}}^H \bm{R}_X^* \dot{\bm{V}}_{\bm{y}} \bm{B}).
\end{align}
We can find $\bm{\mathrm{F}}_{\bm{xz}}$ and $\bm{\mathrm{F}}_{\bm{yz}}$ by changing $\bm{xy}$ in (\ref{equ:Gauss_CRB_xy}) and (\ref{equ:Gauss_CRB_XY}) into $\bm{xz}$ and $\bm{yz}$, respectively, and accordingly find $\bm{\mathrm{F}}(\bm{x},\bm{z}) = 2 \mathfrak{R}(\bm{\mathrm{F}}_{\bm{xz}})$ and $\bm{\mathrm{F}}(\bm{y},\bm{z}) = 2 \mathfrak{R}(\bm{\mathrm{F}}_{\bm{yz}})$.
As the FIM is inherently a symmetric positive semidefinite matrix, we accordingly have $\bm{\mathrm{F}}(\bm{y},\bm{x}) = \bm{\mathrm{F}}^T(\bm{x},\bm{y})$, $\bm{\mathrm{F}}(\bm{z},\bm{x}) = \bm{\mathrm{F}}^T(\bm{x},\bm{z})$, and $\bm{\mathrm{F}}(\bm{z},\bm{y}) = \bm{\mathrm{F}}^T(\bm{y},\bm{z})$.


Third, we find $\bm{\mathrm{F}}(\bm{b}_\text{R},\bm{b}_\text{R})$, $\bm{\mathrm{F}}(\bm{b}_\text{I},\bm{b}_\text{I})$, and $\bm{\mathrm{F}}(\bm{b}_\text{R}, \bm{b}_\text{I})$. Towards this end, we have
\begin{align}\label{equ:Gauss_CRB_b_RI_general1}
\bm{\mathrm{F}}(b_{\text{R}_i},b_{\text{R}_j}) &= 2 \mathfrak{R}  \operatorname{tr} \left[ (\frac{\partial (\bm{A} \bm{B} \bm{V}^T \bm{X})}{\partial b_{\text{R}_i}})^H \bm{Q}^{-1} (\frac{\partial (\bm{A} \bm{B} \bm{V}^T \bm{X})}{\partial b_{\text{R}_j}}) \right],
\end{align}
\begin{align}\label{equ:Gauss_CRB_b_RI_general2}
\bm{\mathrm{F}}(b_{\text{I}_i},b_{\text{I}_j}) &= 2 \mathfrak{R}  \operatorname{tr} \left[(\frac{\partial (\bm{A} \bm{B} \bm{V}^T \bm{X})}{\partial b_{\text{I}_i}})^H \bm{Q}^{-1} (\frac{\partial (\bm{A} \bm{B} \bm{V}^T \bm{X})}{\partial b_{\text{I}_j}}) \right],
\end{align}
\begin{align}\label{equ:Gauss_CRB_b_RI_general3}
\bm{\mathrm{F}}(b_{\text{R}_i},b_{\text{I}_j}) &= 2 \mathfrak{R}  \operatorname{tr} \left[(\frac{\partial (\bm{A} \bm{B} \bm{V}^T \bm{X})}{\partial b_{\text{R}_i}})^H \bm{Q}^{-1} (\frac{\partial (\bm{A} \bm{B} \bm{V}^T \bm{X})}{\partial b_{\text{I}_j}}) \right],
\end{align}
and 
\begin{align}\label{equ:Gauss_CRB_b_Real}
\frac{\partial\left(\bm{A B} \bm{V}^T \boldsymbol{X}\right)}{\partial b_{\text{R}_i}}&=\bm{A} \bm{e}_i \bm{e}_i^T \bm{V}^T \boldsymbol{X}, \\
\label{equ:Gauss_CRB_b_Imag}
\frac{\partial\left(\bm{A B} \bm{V}^T \boldsymbol{X}\right)}{\partial b_{\text{I}_i}}&= \mathrm{j} \bm{A} \bm{e}_i \bm{e}_i^T \bm{V}^T \boldsymbol{X}.
\end{align}
Thus, similarly, we have $\bm{\mathrm{F}}(\bm{b}_\text{R},\bm{b}_\text{R}) = 2 \mathfrak{R}(\bm{\mathrm{F}}_{\bm{bb}})$, where
\begin{align}\label{equ:Gauss_CRB_b_R}
\bm{\mathrm{F}}_{\bm{bb}} = L (\bm{A}^H  \bm{Q}^{-1} \bm{A}) \odot (\bm{V}^H \bm{R}_X^* \bm{V}), 
\end{align}
$\bm{\mathrm{F}}(\bm{b}_\text{I},\bm{b}_\text{I}) = 2 \mathfrak{R}(\bm{\mathrm{F}}_{\bm{bb}})$, and $\bm{\mathrm{F}}(\bm{b}_\text{R},\bm{b}_\text{I}) = - 2 \mathfrak{I}(\bm{\mathrm{F}}_{\bm{bb}})$
Accordingly, we have $\bm{\mathrm{F}}(\bm{b}_\text{I},\bm{b}_\text{R}) = \bm{\mathrm{F}}^T(\bm{b}_\text{R},\bm{b}_\text{I}) = - 2 \mathfrak{I}(\bm{\mathrm{F}}^T_{\bm{bb}})$.

Finally, we find $\bm{\mathrm{F}}(\bm{x},\bm{b}_\text{R})$ and $\bm{\mathrm{F}}(\bm{x},\bm{b}_\text{I})$. We then have 
\begin{align}
\nonumber
\bm{\mathrm{F}}(x_i,b_{\text{R}_j}) & = 2 \mathfrak{R} \operatorname{tr} \left[ (\frac{\partial (\bm{A} \bm{B} \bm{V}^T \bm{X})}{\partial x_i})^H \bm{Q}^{-1} (\frac{\partial (\bm{A} \bm{B} \bm{V}^T \bm{X})}{\partial b_{\text{R}_j}}) \right] \\
\nonumber
& = 2 \mathfrak{R} \operatorname{tr} \left[ \left(\dot{\bm{A}_{\bm{x}}} \bm{e}_i \bm{e}_i^T \bm{B} \bm{V}^T \boldsymbol{X}+\bm{A B} \bm{e}_i \bm{e}_i^T \dot{\bm{V}}_{\bm{x}}^T \boldsymbol{X}\right)^H \right.\\
& \left. \bm{Q}^{-1} \left(\bm{A} \bm{e}_j \bm{e}_j^T \bm{V}^T \boldsymbol{X}\right) \right].
\end{align}
Similarly, one can obtain $\bm{\mathrm{F}}(\bm{x},\bm{b}_\text{R}) = \bm{\mathrm{F}}^T(\bm{b}_\text{R},\bm{x}) = 2 \mathfrak{R}(\bm{\mathrm{F}}_{\bm{xb}})$, where
\begin{align}\label{equ:Gauss_CRB_b_xR}
\nonumber
\bm{\mathrm{F}}_{\bm{xb}} & = L (\dot{\bm{A}}_{\bm{x}}^H  \bm{Q}^{-1} \bm{A}) \odot (\bm{B}^* \bm{V}^H \bm{R}_X^* \bm{V})\\
& + L (\bm{A}^H  \bm{Q}^{-1} \bm{A}) \odot (\bm{B}^* \dot{\bm{V}}_{\bm{x}}^H \bm{R}_X^* \bm{V}).
\end{align}
We can find $\bm{\mathrm{F}}_{\bm{yb}}$ and $\bm{\mathrm{F}}_{\bm{zb}}$ by changing $\bm{x}$ in (\ref{equ:Gauss_CRB_b_xR}) into $\bm{y}$ and $\bm{z}$, respectively. Accordingly, we have $\bm{\mathrm{F}}(\bm{y},\bm{b}_\text{R}) = \bm{\mathrm{F}}^T(\bm{b}_\text{R},\bm{y}) = 2 \mathfrak{R}(\bm{\mathrm{F}}_{\bm{yb}})$ and $\bm{\mathrm{F}}(\bm{z},\bm{b}_\text{R}) = \bm{\mathrm{F}}^T(\bm{b}_\text{R},\bm{z}) = 2 \mathfrak{R}(\bm{\mathrm{F}}_{\bm{zb}})$.
%

For $\bm{\mathrm{F}}(\bm{x},\bm{b}_\text{I})$, $\bm{\mathrm{F}}(\bm{y},\bm{b}_\text{I})$, and $\bm{\mathrm{F}}(\bm{z},\bm{b}_\text{I})$, according to  (\ref{equ:Gauss_CRB_b_Imag}), we similarly have 
\begin{align}
\bm{\mathrm{F}}(\bm{x},\bm{b}_\text{I}) = \bm{\mathrm{F}}^T(\bm{b}_\text{I},\bm{x}) = -2 \mathfrak{I}(\bm{\mathrm{F}}_{\bm{xb}}),
\end{align}
\begin{align}
\bm{\mathrm{F}}(\bm{y},\bm{b}_\text{I}) = \bm{\mathrm{F}}^T(\bm{b}_\text{I},\bm{y}) = -2 \mathfrak{I}(\bm{\mathrm{F}}_{\bm{yb}}),
\end{align}
and 
\begin{align}
\bm{\mathrm{F}}(\bm{z},\bm{b}_\text{I}) = \bm{\mathrm{F}}^T(\bm{b}_\text{I},\bm{z}) = -2 \mathfrak{I}(\bm{\mathrm{F}}_{\bm{zb}}).
\end{align}
Combining the above yields the proof.

\bibliographystyle{ieeetran}

\bibliography{refsv2}

\end{document}